\documentclass{article}
\begin{document}

\title{Group classification of the general evolution equation: local
and quasilocal symmetries}

\author{ Renat ZHDANOV \thanks{Email:~renat@imath.kiev.ua} \\
\small Institute of Mathematics, 3 Tereshchenkivska Street,\\
\small Kyiv 4, 01601, Ukraine\\ \\
Victor LAHNO \thanks{Email:~laggo@poltava.bank.gov.ua}\\
\small State Pedagogical University, 2 Ostrogradskogo Street,\\
\small Poltava 36003, Ukraine}

\date{}

\maketitle

\begin{abstract}
We give a review of our recent results on group classification of
the most general nonlinear evolution equation in one spatial
variable. The method applied relies heavily on the results of our
paper {\em Acta Appl. Math.}, {\bf 69}, 2001, in which we obtain the
complete solution of group classification problem for general
quasilinear evolution equation.
\end{abstract}

In this paper we briefly review our recent results on group
classification of the general nonlinear evolution equation
\begin{equation}
 u_t=F(t,x,u,u_x,u_{xx}). \label{1}
\end{equation}
Here $u=u(t,x)$,\ $u_t={\partial u}/{\partial t}$,\ $u_x = {\partial
u}/{\partial x}$, $u_{xx}={\partial^2 u}/{\partial x^2}$;\ $F$ is an
arbitrary smooth function obeying the restriction\ ${\partial
F}/{\partial u_{xx}}\not=0$.

Using the standard Lie approach we prove that the maximal invariance
group of equation (\ref{1}) is generated by the operator
\begin{equation}
{\rm Q} = \tau(t)\partial_t+\xi(t,x,u)\partial_x+
\eta(t,x,u)\partial_u, \label{6}
\end{equation}
where functions $\tau,\xi$ and $\eta$ are arbitrary solutions of a
single partial differential equation (PDE)
\begin{eqnarray}
&&\eta_t-u_x\xi_t+(\eta_u-\tau_t-u_x\xi_u)F
=\Bigl(\eta_x+u_x(\eta_u-\xi_x)-u^2_x\xi_u\Bigr)F_{u_x}\nonumber \\
&&\quad+\Bigl(\eta_{xx}+u_x(2 \eta_{xu}-\xi_{xx})+u^2_x(\eta_{uu}-
2\xi_{xu})-u^3_x\xi_{uu} \label{7} \\
&&\quad+u_{xx}(\eta_u-2\xi_x)-3u_xu_{xx}\xi_u\Bigr)F_{u_{xx}}+\tau
F_t +\xi F_{x}+\eta F_u. \nonumber
\end{eqnarray}

So to obtain (exhaustive) group classification of the class of
equations (\ref{1}) we need to construct {\em all} possible
functions, $\tau, \xi, \eta$ and $F$, obeying the above constraint
(determining equation). Evidently the challenge of the problem is in
the word {\em all}, since the system of classifying equations is not
over-determined (as is customary for this type of problems).
Moreover it is under-determined. This is the reason why the numerous
papers devoted to group classification of nonlinear evolution
equations deal mostly with classes of PDEs depending on arbitrary
functions of one, or at most two, variables.

A starting point of our analysis is a simple observation that
solutions ${\mathbf v}_a = (\tau_a, \xi_a, \eta_a),\ a=1,\ldots,n$,
of (\ref{7}) span a Lie algebra $\ell$. So without any loss of
generality we can replace (\ref{7}) with the (possibly infinite) set
of systems of PDEs
\begin{equation}\nonumber
\left\{
\begin{array}{l}
{\rm Equation (\ref{7})},\\[2mm]
[Q_i,\ Q_j]=C_{ij}^k\,Q_k,
\end{array}
\right.
\end{equation}
or, equivalently,
\begin{equation}
\label{9} \left\{
\begin{array}{l}
{\rm Equation (\ref{7})},\\[2mm]
Q_i\tau_j-Q_j\tau_i=C_{ij}^k\,\tau_k,\\[2mm]
Q_i\xi_j-Q_j\xi_i=C_{ij}^k\,\xi_k,\\[2mm]
Q_i\eta_j-Q_j\eta_i=C_{ij}^k\,\eta_k.
\end{array}
\right.
\end{equation}
In the above formulas the indices $i,j,k$ take the values
$1,\ldots,n$ ($n\ge 1$ is a dimension of the corresponding Lie
algebra), and $C_{ij}^k$ are structure constants of the Lie algebra
$\ell$.

If we solve the (over-determined) system of PDEs (\ref{9}) for {\em
all} possible dimensions $n\ge 1$ of {\em all} admissible Lie
algebras, $\ell$, then the problem of group classification of
Eq.(\ref{1}) is completely solved. In other words the problem of
group classification of the general evolution equation (\ref{1})
reduces to integrating over-determined systems of PDEs (\ref{9}) for
all $n=1,2,\ldots,n_0$, where $n_0$ is the maximal dimension of the
Lie algebra admitted by the equation under study.

One way to handle the above problem would be starting with
investigating compatibility of systems (\ref{9}) for all $n\ge 1$.
This strategy is close in spirit to Reid's procedure of describing
the algebra admitted by PDE without integrating determining
equations \cite{reid}.

However, a more natural approach is actually to integrate equations
(\ref{9}) so that compatibility conditions come as a by-product.
This is even more so if we take into account that low-dimensional
abstract Lie algebras are described up to the dimension $n=6$
(mainly due to efforts by Mubarakzyanov \cite{mub1,mub2}). So, if we
\begin{enumerate}
\itemsep=0pt \item construct all realizations of Lie algebras by
operators, the coefficients of which satisfy Eq.(\ref{7}), up to
some fixed dimension $n_0$, and \item prove that (\ref{1}) does not
admit invariance algebras of the dimension $n>n_0$,
\end{enumerate}
then the problem of group classification of Eq.(\ref{1}) is
completely solved.

The underlying ideas of the above approach are rather natural. No
wonder that they have already been used in various contexts. In
particular Fushchych \& Serov \cite{ren00} exploited them to
classify conformally-invariant wave equations in the
multidimensional case. In a more systematic way these ideas have
been utilized by Gagnon \& Winternitz  \cite{ren01} to classify
variable coefficient Schr\"odinger equations.

In its present form the approach formulated above has been developed
in \cite{ren02}, where we perform preliminary group classification
of nonlinear Schr\"odinger equations. Later we applied this approach
to classify second quasilinear evolution equations \cite{ren1,ren2},
third-order evolution equations \cite{ren4} and nonlinear wave
equations \cite{ren5} in one spatial variable.

We perform group classification within the action of equivalence
group preserving the class of PDEs under study. It is not difficult
to prove that the maximal equivalence transformation group
preserving class (\ref{1}) is
\begin{equation}
\bar t=T(t),\quad \bar x=X(t,x,u),\quad \bar u=U(t,x,u), \label{8}
\end{equation}
where
\begin{equation}
\nonumber T'=\frac{dT}{dt}\not=0,\quad
\frac{D(X,U)}{D(x,u)}\not=0.
\end{equation}

In the paper \cite{ren1} we obtain an exhaustive group
classification of (quasilinear) evolution equation
\begin{equation}
v_t=f(t,x,v,v_x)v_{xx}+g(t,x,v,v_x),\quad v=v(t,x). \label{2}
\end{equation}
Those results provide almost complete solution of the problem of
group classification for general evolution equation (\ref{1}). This
claim follows from the fact that, if Eq.(\ref{1}) admits a
one-parameter group with the infinitesimal generator
\begin{equation}
Q=\xi(t,x,u)\partial_x+\eta(t,x,u)\partial_u, \label{3}
\end{equation}
then it is transformed into an equation of the form (\ref{1}).

Indeed operator (\ref{3}) can be reduced to the canonical form
$\partial_{u^\prime}$ by a suitable change of variables
\begin{equation}
 t^{\prime}=T(t),\quad x^{\prime}=X(t,x,u),\quad \label{4}
u^{\prime}=U(t,x,u)
\end{equation}
(note that the above transformation belongs to the equivalence group
of equation (\ref{1})). The corresponding invariant equation takes
the form
$$
u^{\prime}_{t^{\prime}} =
F^{\prime}(t^{\prime},x^{\prime},u^{\prime}_{x^{\prime}},
u^{\prime}_{x^{\prime}x^{\prime}}).
$$

Differentiating the obtained equation by $x^{\prime}$, replacing
$u^{\prime}_{x^{\prime}}$ with $v(t^{\prime},x^{\prime})$ and
dropping the primes we arrive at the quasilinear PDE of the form
(\ref{2}).

Before proceeding to exploit the above fact any further, we
briefly summarize the principal results of \cite{ren1}.
\begin{itemize}
\itemsep=0pt
\item There are 2 inequivalent PDEs (\ref{2}) admitting
one-dimensional algebras;

\item There are 5 inequivalent PDEs (\ref{2}) admitting
two-dimen\-si\-on\-al algebras.

\item There are 34 inequivalent PDEs (\ref{2}) admitting
three-di\-men\-si\-on\-al algebras.

\item There are 35 inequivalent PDEs (\ref{2}) admitting
four-dimen\-si\-on\-al algebras.

\item There are 6 inequivalent PDEs (\ref{2}) admitting
five-dimen\-si\-on\-al algebras.
\end{itemize}

As an example, we give below the complete list of inequivalent
equations (\ref{2}) invariant under five-dimensional Lie algebras.
\begin{eqnarray}
&&u_t=u^{-4}\,u_{xx}-2u^{-5}\,u_x^2,\nonumber\\
&&u_t=u_{xx}+x^{-1}\,u\,u_x-x^{-2}\,u^2-2x^{-2}\,u,\nonumber\\
&&u_t=u_x^{-2}\,u_{xx}+u_x^{-1},\nonumber\\
&&u_t={\rm e}^{u_x}\,u_{xx},\nonumber\\
&&u_t=u_x^n\,u_{xx},\quad n\ge -1,\ n\not=0,\nonumber\\
&&u_t=(1+u_x^2)^{-1}\,\exp(n\arctan u_x)\,u_{xx}.\nonumber
\end{eqnarray}

It follows from the above considerations that, if the invariance
algebra of equation (\ref{1}) contains an operator of the form
(\ref{3}), then it is equivalent to equation (\ref{2}) the group
properties of which are already known so that to complete group
classification of Eq.(\ref{1}) we need to describe all equations
(\ref{1}) the invariance algebras of which are spanned by the
operators
\begin{equation}
 Q_i =
\tau_i(t)\partial_t+\xi_i(t,x,u)\partial_x+\eta_i(t,x,u)\partial_u,\quad
i=1,\ldots,n, \label{5}
\end{equation}
where the functions $\tau_1(t),\ldots,\tau_n(t)$ are linearly
independent. We denote the class of such equations as ${\cal
L}_1$.

We prove that the highest dimension of an invariance algebra of
Eq.(\ref{2}) belonging to ${\cal L}_1$ equals 3. The algebra in
question is $sl(2,\textbf{R})$ and the corresponding invariant
equations are given below.
\begin{eqnarray}
&&u_t=x^{-1}\,u\,u_x-x^{-2}\,u^2+x^{-2}\,\tilde F(x^2u_{xx}-2u,\ 2u-xu_x), \nonumber\\
&&\quad  sl(2,\textbf{R})=\left\langle 2t\partial_t+x\partial_x,\
-t^2\partial_t-tx\partial_x+x^2\partial_u,\
\partial_t\right\rangle; \nonumber\\
&&u_t=-\frac{1}{4}\,x^{-1}\,u_x+x^{-3}\,u^{-1}_x\,\tilde F(u,\
u^{-2}_xu_{xx}+3x^{-1}u^{-1}_x); \nonumber\\
&&\quad  sl(2,\textbf{R})=\left\langle 2t\partial_t+x\partial_x,\
-t^2\partial_t+x(x^2-t)\partial_x,\
\partial_t\right\rangle. \nonumber
\end{eqnarray}
Here $\tilde F$ is an arbitrary smooth function.

There are only two equations from ${\cal L}_1$ admitting
lower-dimensional invariance algebras, namely,
\begin{eqnarray}
&&u_t=\tilde F(x,\ u^{-1}u_x,\ u^{-1}u_{xx}),\quad \ell=\left\langle
-t\partial_t-u\partial_u,\ \partial_t\right\rangle. \nonumber\\
&&u_t=\tilde F(x,\ u,\ u_x,\ u_{xx}),\quad \ell=\left\langle
\partial_t\right\rangle. \nonumber
\end{eqnarray}
In the above formulas $\tilde F$ is an arbitrary smooth function.

Equations from ${\cal L}_1$ together with invariant equations of the
form (\ref{2}) provide the complete solution of the problem of
classifying equations (\ref{1}) that admit nontrivial Lie symmetry.

As we noted in \cite{ren2}, results of the group classification of
(\ref{1}) can be utilized to derive their quasilocal symmetries. The
term quasi-local has been introduced independently in \cite{ren3}
and \cite{pukh} to distinguish nonlocal symmetries that are
equivalent to local ones through nonlocal transformation.

We have already shown that equations (\ref{1}) and (\ref{2}) are
related through the nonpoint transformation $v(t,x)=u_x(t,x)$ or,
inversely, $u(t,x)=\partial_x^{-1}v(t,x)$. Suppose now that
Eq.(\ref{1}) admits the one-parameter transformation group
$$
\left\{
\begin{array}{l}
t^\prime=T(t,\theta),\\[1mm]
x^\prime=X(t,x,u,\theta),\\[1mm]
u^\prime=U(t,x,u,\theta).
\end{array}
\right.
$$
Computing the first prolongation of the above formulas gives the
transformation rule for the first derivative of $u^\prime$
$$
\frac{\partial u^\prime}{\partial x^\prime} =
\frac{u_xU_u+U_x}{u_xX_u+X_x}.
$$
In terms of the variables, $t,x,v(t,x)$, this transformation group
is
$$
\left\{
\begin{array}{l}
t^\prime=T(t,\ \theta),\\[1mm]
x^\prime=X(t,\ x,\ u,\ \theta),\\[3mm]
v^\prime= \frac{\mbox{$vU_u(t,\ x,\ u(t,\,x),\ \theta)+U_x(t,\ x,\
u(t,\,x),\ \theta)$}}{\mbox{$vX_u(t,\ x,u(t,\,x),\ \theta)+X_x(t,\
x,\ u(t,\,x),\ \theta)$}},
\end{array}
\right.
$$
where $u(t,x)=\partial^{-1}v(t,x)$. Consequently, if the relation
\begin{equation}
X_{uu}^2+X_{xu}^2+U_{uu}^2+U_{ux}^2\not=0 \label{10}
\end{equation}
holds, the transformed equation (\ref{2}) possesses a quasilocal
symmetry. If a symmetry group of Eq.(\ref{1}) satisfies constraint
(\ref{10}), then we say that this equation belongs to the class
${\cal L}_2$. In what follows we describe all equations from ${\cal
L}_2$, the symmetry algebras of which are at most three-dimensional.

It is not difficult to become convinced of the fact that the class
${\cal L}_2$ does not contain equations the maximal invariance
algebras of which are of the dimension $n\le 2$. Below we give the
full list of inequivalent equations belonging to ${\cal L}_2$ and
admitting three-dimensional Lie algebras (we follow notations of
\cite{ren2})
\begin{eqnarray*}
&&\textbf{Algebra}\ sl(2,\textbf{R}):\\
&1.&{\rm Realization}\\
&&\quad Q_1=\partial_u, \ \ \quad Q_2=2u\partial_u-x\partial_x,\ \
\quad Q_3=-u^2\partial_u+xu \partial_x.\\
&&\quad {\rm Invariant\ equation:} \\
&&\quad u_t=xu_x\,\tilde F(t,\ x^{-5}\,u^{-3}_x\,u_{xx}+2 x^{-6}\,
u^{-2}_x). \\
&2.&{\rm Realization} \\
&&\quad Q_1=\partial_u, \ \ \quad Q_2=2u\partial_u-x\partial_x. \\
&&\quad Q_3=(\varepsilon x^{-4}-u^2)\partial_u+xu \partial_x, \ \varepsilon=\pm 1. \\
&&\quad {\rm Invariant\ equation:} \\
&&\quad u_t=x^{-2}\,\sqrt{x^6\,u^2_x+4\varepsilon}\,\tilde
F\left(t,\ (x^6\,u^2_x+4\varepsilon)^{-\frac{3}{2}}\,(x^4\,u_{xx}
+5x^3\,u_x+\frac{1}{2}\,x^9\,u^3_x )\right).\\
&&\textbf{Algebra}\ so(3):\\
&1.& {\rm Realization}  \\
&&\quad Q_1=\partial_u,  \ \ \quad Q_2=\cos u\partial_x+\tan x\sin
u\partial_u,  \\
&&\quad Q_3=-\sin u\partial_x+\tan x\cos u\partial_u.  \\
&&\quad {\rm Invariant\ equation:}  \\
&&\quad u_t=\sqrt{\sec^2 x+u^2_x}\,\tilde F\left(t,\ \left(u_{xx}\,
\cos x-(2+u^2_x\,\cos^{2}x)\,u_x\,\sin x\right)\right.\\
&&\quad\times\left.(1+u^2_x\,\cos^{2}x)^{-\frac{3}{2}}\right).\\
&&\textbf{Algebra}\ A_{3.8}:\\
&1.&{\rm Realization}  \\
&&\quad Q_1=\partial_u,  \ \ \quad Q_2=x\partial_u,  \ \
\quad Q_3=-(x^2+1)\partial_x-xu\partial_u.  \\
&&\quad {\rm Invariant\ equation}  \\
&&\quad u_t=\sqrt{1+x^2}\,\tilde
F\left(t,\ u_{xx}\,(1+x^2)^{\frac{3}{2}}\right).  \\
&2.&{\rm Realization}  \\
&&\quad Q_1=\partial_u,  \ \ \quad Q_2=-\tan(t+x)\partial_u,  \ \
\quad Q_3=\partial_t+\tan(t+x)u\partial_u.  \\
&&\quad {\rm Invariant\ equation}  \\
&&\quad u_t=u_x+\sec(t+x)\,F\left(x,\
u_{xx}\,\cos(t+x)-2u_x\,\sin(t+x)\right).\\
&&\textbf{Algebra}\ A_{3.9}:\\
&1.&{\rm Realization}  \\
&&\quad Q_1=\partial_u,  \ \
\quad Q_2=x\partial_u,  \\
&&\quad Q_3=-(x^2+1)\partial_x+(q - x)u\partial_u,  \ q\not =0.\\
&&\quad {\rm Invariant\ equation}  \\
&&\quad u_t={\rm e}^{-q\arctan x}\,\sqrt{1+x^2}\,\tilde F\left(t,\
u_{xx}\,{\rm e}^{q\arctan x}\,(1+x^2)^{\frac{3}{2}}\right).\\
&2.&{\rm Realization}  \\
&&\quad Q_1=\partial_u,  \ \
\quad Q_2=-\tan(t+x)\partial_u,  \\
&&\quad Q_3=\partial_t+(q+\tan(t+x))u\partial_u, \ q\not =0.\\
&&\quad {\rm Invariant\ equation}  \\
&&\quad u_t=u_x+\sec(t+x)\,{\rm e}^{qt}\,\tilde F\left(x,\ {\rm
e}^{-qt}\,(u_{xx}\,\cos(t+x)-2u_x\,\sin(t+x))\right).
\end{eqnarray*}

Differentiating any of the above equations by $x$ and replacing
$u_x$ with $v$ yields an equation of the form (\ref{2}) that admits
a quasilocal symmetry. Consider, as an example, equation
$$
u_t=u_x+\sec(t+x)\,F\left(x,\
u_{xx}\,\cos(t+x)-2u_x\,\sin(t+x)\right)
$$
which is invariant with respect to the algebra $\langle
\partial_t+\tan(t+x)u\partial_u \rangle$. The corresponding
one-para\-me\-ter transformation group is
$$
\left\{
\begin{array}{l}
t^\prime=t,\\
x^\prime=x,\\
u^\prime=u\,\sec(t+x+\theta),
\end{array}
\right.
$$
where $\theta\in\mathbf{R}$ is the group parameter. Computation of
the first prolongation of the above formulas yields
$$
u^\prime_{x^\prime}=u_x\,\sec(t+x+\theta)+
u\,\sec(t+x+\theta)\,\tan(t+x+\theta)
$$
or
$$
v^\prime_{x^\prime}(t^\prime,\,x^\prime)=\Bigl(v(t,\,x)+
u(t,\,x)\,\tan(t+x+\theta)\Bigr)\sec(t+x+\theta),
$$
where $u(t,x)=\partial^{-1}v(t,x)$. The corresponding equation for
$v=v(t,x)$ is
$$
v_t=v_x+\sec(t+x)\ (\tan(t+x)\ \tilde F+\tilde F_{\omega_1})+
(v_{xx}-3\tan(t+x)\,v_x-2v)\ \tilde F_{\omega_2}.
$$
Here $\tilde F$ is an arbitrary smooth function of the
variables
$$
\omega_1=x\ {\rm and}\ \omega_2=\cos(t+x)\,v_x-2\sin(t+x)\,v.
$$

\section{Concluding Remarks}
The motivation for writing this paper is to introduce our strategy
for attacking the problem of group classification of the most
general evolution equation in one spatial dimension. The three basic
elements of our analysis are
\begin{itemize}
\item Group classification of the partial case of the class of PDEs
in question, of the quasilinear evolution equations.
\item Description of nonlinear PDEs, which belong to the class
${\cal L}_1$.
\item Description of nonlinear PDEs, which belong to the class
${\cal L}_2$.
\end{itemize}
It is one of the principal results of the present paper that the
complete solution of these three subproblems yields a final solution
of the problem of group classification of the class of PDEs
(\ref{1}).

While the first two subproblems have already been solved (see,
cite{ren1,ren2} and this paper), the classification of ${\cal L}_2$
evolution equations is still to be completed. We present the
classification results for the Lie algebras which have dimension not
higher than three. The work on higher dimensional Lie algebras is in
progress now and will be reported elsewhere.

\end{document}